\documentclass[aps,prd,preprint,groupedaddress,nofootinbib,amsmath]{revtex4-1}

\usepackage{amsmath}
\usepackage{amssymb}
\usepackage{physics}
\usepackage{bbold}
\usepackage{graphicx}
\usepackage{slashbox}
\usepackage{verbatim}

\usepackage{hyperref}

\usepackage{indentfirst}

\newcommand{\be}{\begin{equation}}
\newcommand{\en}{\end{equation}}
\newcommand{\bea}{\begin{align}}
\newcommand{\ena}{\end{align}}

\begin{document}

\title{Fermion Production in Bouncing Cosmologies}
\author{A. Scardua}\email{arthur@cbpf.br}
\affiliation{CBPF - Centro Brasileiro de
Pesquisas F\'{\i}sicas, Xavier Sigaud st. 150,
zip 22290-180, Rio de Janeiro, Brazil.}

\author{L.~F.~Guimar\~aes}\email{lfog@cbpf.br}
\affiliation{CBPF - Centro Brasileiro de
Pesquisas F\'{\i}sicas, Xavier Sigaud st. 150,
zip 22290-180, Rio de Janeiro, Brazil.}

\author{N.~Pinto-Neto}\email{nelson.pinto@pq.cnpq.br}
\affiliation{CBPF - Centro Brasileiro de
	Pesquisas F\'{\i}sicas, Xavier Sigaud st. 150,
	zip 22290-180, Rio de Janeiro, Brazil.}

\author{G.~S.~Vicente}\email{gustavosvicente@gmail.com}
\affiliation{CBPF - Centro Brasileiro de
	Pesquisas F\'{\i}sicas, Xavier Sigaud st. 150,
	zip 22290-180, Rio de Janeiro, Brazil.}

\date{\today}

\begin{abstract}
	We address the issue of fermionic particle creation in cosmological bouncing models governed by General Relativity, but where the bounce itself takes place due to quantum effects. If the energy scale of the bounce is not very close to the Planck energy, the Wheeler-DeWitt approach can be used to furnish sensible singularity-free background models with a contracting phase preceding an expanding phase in which the fermions evolve. The fermionic fields considered are massive, neutral and minimally coupled to gravity. We are particularly interested in neutrinos, neutrons and heavy neutrinos motivated by theories beyond the Standard Model of Particle Physics. We perform a numerical analysis for a bouncing model containing radiation and a pressureless fluid. The results reveal that the fermionic production is very small, with no back-reaction, unless the fermions are very heavy with masses up to $10^9$GeV. Hence, investigations concerning baryogenesis in such bouncing models should either go beyond the minimal coupling between gravity and the fermionic fields considered here, or assume the existence of such heavy fermions as a starting point.
\end{abstract}


\maketitle

\section{Introduction}
In a dynamical Universe, in which there is no time-like killing vector field, particles can be created by the gravitational field. This is usually done through parametric amplification, which is evoked by many authors as the process of creation of particles in the pre-heating phase of inflationary models~\cite{Kolb:1990vq, Lyth:1998xn,Riotto,Peloso:2000hy}, and as the amplification process of cosmological perturbations originated from quantum vacuum fluctuations~\cite{Mukhanov:2005sc}. 

In cosmological bouncing scenarios, parametric amplification is also responsible for the enhancement of cosmological perturbations along its evolution~\cite{Finelli:2001sr,Peter:2006hx,Peter:2008qz,Peter:2015zaa,Vitenti:2011yc}, and for the creation of scalar particles during the bounce~\citep{Celani:2016cwm}. In both cases, an initial vacuum state is defined through an adiabatic prescription when the Universe was very big and almost flat in the far past of the contracting phase, and the time dependent gravitational field acts as a pump field along the cosmological evolution. In the case of cosmological perturbations, amplitudes and spectra of scalar and tensor perturbations can be obtained compatible with the observed values, provided the contracting phase is dominated by a pressureless fluid (compatible with dark matter)~\citep{Wands:1998yp,Bacalhau:2017hja}. Non-gaussianities of such perturbations are now being investigated~\citep{Quintin:2015rta}. In the case of scalar particle creation, the production is usually small, although in some special cases it can be comparable to the background energy density, and back-reaction must be evaluated~\citep{Celani:2016cwm}.

The aim of this paper is to investigate fermion creation in bouncing models which some of us have been investigating along the past decades. In these models, the bounce occurs due to quantum cosmological effects when the curvature of space-time becomes very large, up to scales a few orders of magnitude below the Planck energy (for more involved theories suitable for energy scales close to or above the Planck scale, see Refs.~\cite{Ashtekar1,Ashtekar2} as some examples). In this case, the Wheeler-DeWitt approach is reliable. Note that the usual Copenhagen point of view cannot be used in quantum cosmology (see Ref.~\cite{Pinto-Neto:2013toa} for a review on this subject), hence we interpreted the solutions using the de Broglie-Bohm quantum theory~\cite{HollandBook}. In this framework, trajectories can be defined, and the scale factor evolution can be calculated. They are usually nonsingular, presenting a bounce due to quantum effects at small scales, and turning to a classical standard evolution when the scale factor becomes sufficiently large~\cite{Peter:2006hx,Pinto-Neto:2013toa, AcaciodeBarros:1997gy, Alvarenga:2001nm}. The models considered in this paper contain one single hydrodynamical fluid or two fluids~\cite{PintoNeto:2005gx}: the usual observed radiation and dust contents which are present in our universe.

The fermionic fields we consider in this paper are massive neutral fermions minimally coupled to gravity in the standard way. The fermions considered are the neutrino, the neutron, and other massive fermions beyond the Standard Model of Particle Physics with masses up to order $m = 10^9$GeV. For masses far below $10^9$GeV, the energy density of created fermions is much smaller than the background energy density, even for bounce energies as close as possible to the Planck energy. Hence, no back-reaction will arise. For masses of this order, however, the energy density of created fermions can be comparable to the background energy density, and back-reaction should be considered. Hence, if one wants to consider baryogenesis~\cite{RiottoTrodden} in bouncing models, these heavy fermions beyond the Standard Model should be assumed to exist in the early universe and then decayed to the standard fermions we know. Another possibility is to consider new different couplings between the fermionic fields and gravity (see Refs.~\citep{Antunes:2016efv,Davoudiasl:2004gf} as examples), not only to yield an overproduction of particles on anti-particles, but also to enhance the fermionic production with respect to the standard minimal coupling case.

This paper is divided as follows: in Sec.~\ref{sec2} we review in some detail the standard theoretical derivation of fermion production in a time-dependent homogeneous and isotropic space-time. In Sec.~\ref{sec3}, we shortly review the properties of the bouncing cosmological backgrounds we will consider, which contains one or two background fluids (radiation and a pressureless fluid), and we apply the results of Sec.~\ref{sec2} to these situations. We obtain the expressions for the created number density in terms of the Bogoliubov coefficients, the fermion masses, and the background parameters. In Sec.~\ref{sec4}, we perform the numerical integration to obtain the number density of created fermions, and we discuss their physical impact. We conclude in Sec.~\ref{sec5} with a summary of our results, and a discussion of future developments. In Appendix~\ref{appendix:A}, we study the infrared and ultra-violet limits in order to compare them with the numerical calculations.

\section{Fermion Creation in Curved Spaces}
\label{sec2}

In this section, we present the formalism of Dirac fermion creation in curved space-times, following Refs.~\cite{Riotto,Peloso:2000hy}. 
We consider the case of a spatially flat Friedmann-Lema\^{\i}tre-Robertson-Walker (FLRW) space-time, whose line element is given by ${\rm d} s^2 = a^2(\eta)({\rm d}\eta^2-{\rm d}{{\vec{x}}^2)}$, where $a$ is the scale factor and $\eta$ is conformal time.
The Dirac equation in the FLRW space-time reads~\cite{Riotto}
\begin{equation}\label{eq:Dirac_equation}
\left(\frac{i}{a} \gamma^\mu \partial_\mu + i \frac{3}{2} H \gamma^0- m \right) \hat\Psi = 0 ,
\end{equation}
where $H=a'/a^2$ is the Hubble rate, $m$ is the mass of the particle, the prime $'$ represents derivative with respect to conformal time, the $\gamma^{\mu}$ are the $\gamma$-matrices in flat space-time and $\Psi$ is the usual Dirac spinor operator.

Performing the change of variable $\hat\chi \equiv a^{-3/2}\hat\Psi$ for the Dirac spinor, the Dirac equation, Eq.~(\ref{eq:Dirac_equation}), now reads
\begin{equation}\label{eq:Dirac_spin_operator}
\left[i \gamma^\mu \partial_\mu - m a(\eta)\right] \hat\chi = 0  .
\end{equation}
We can work in the momentum representation by taking the Fourier transform of $\hat \chi (\vec x , \eta)$:
\begin{equation}\label{momentum-rep}
\hat \chi (\vec x , \eta) = \frac{1}{(2 \pi)^{3/2}}\int d^3 \vec{k} e^{-i \vec k \cdot \vec x}\hat{\chi}(\vec k , \eta) ,
\end{equation}
where, from Eq.~(\ref{eq:Dirac_equation}), $\hat{\chi}(\vec k , \eta)$ satisfies the equation
\begin{equation}\label{eq-momentum-rep}
\left[i \gamma^0 \partial_\eta + \vec \gamma \cdot \vec k - m a(\eta)\right]\hat{\chi}(\vec k , \eta) =0 .
\end{equation}
Multiplying the latter equation by the unitary operator ${\hat U}_R$, representing a rotation which takes $\vec k$ to the $z$-direction in momentum space yields, after some algebra we obtain
\begin{equation}\label{eq-momentum-rep-z}
\left[i \gamma^0 \partial_\eta + \gamma^3 k - m a(\eta)\right]{\hat{\chi}}_z(\vec k , \eta) =0 ,
\end{equation}
where ${\hat{\chi}}_z \equiv {\hat U}_R \hat{\chi}$ and we have used that ${\hat U}_R \vec \gamma {\hat U}_R = R^i_j \gamma^j = \gamma^3$.

We can expand the Dirac spinor operator ${\hat{\chi}}_z(\vec k , \eta)$ on the basis $S_{l,m}$ of the Dirac spinor space constituted by the simultaneous eigenvectors of $\gamma^0$ and
$\Sigma_z = -\gamma_0 \gamma_3 \gamma 5$, where the latter is proportional to the helicity operator 
$\textup{h}  \equiv \hbar \vec \Sigma \cdot \hat{k}/2 = \hbar \Sigma_z /2$. The basis indexes $l,j = \pm 1$ designate the eigenvalues of $\gamma^0$ and $\Sigma_z$, which in flat space-time discriminate particles from anti-particles and their spin directions, respectively. 
In the standard representation\footnote{
	In the standard representation, the Dirac matrices are given by
	$\gamma^0=\begin{pmatrix}
	\mathbb{1} &0 \\ 
	0 & -\mathbb{1}
	\end{pmatrix}$, \ $\gamma^i=\begin{pmatrix}
	0 &\sigma^i \\ 
	-\sigma^i & 0
	\end{pmatrix}$, \ $\gamma 5=\begin{pmatrix}
	0 &\mathbb{1} \\ 
	\mathbb{1}&0
	\end{pmatrix}$. The $\sigma^i$ are the Pauli matrices. It is also noted that $\gamma_\mu = \eta_{\mu\nu}\gamma^\nu = \left(\gamma^0,-\gamma^i\right)$, where $\eta_{\mu\nu}$ is the Minkowski metric with negative signature.
}, they are given by
\begin{align*}
S_{1,1} =  \begin{pmatrix}1\\0\\0\\0\end{pmatrix}, \quad S_{-1,1} =  \begin{pmatrix}0\\0\\1\\0\end{pmatrix}, \quad
S_{1,-1} =  \begin{pmatrix}0\\1\\0\\0\end{pmatrix}, \quad S_{-1,-1} =  \begin{pmatrix}0\\0\\0\\-1\end{pmatrix}.
\end{align*}
In terms of this basis, the expansion reads
\begin{equation}\label{expansion-basis}
{\hat{\chi}}_z(\vec k , \eta) = \sum_{l,j} S_{l,j} r_{l,j}(\eta, k) {\hat{O}}_{l,j} (k),
\end{equation}
where $r_{l,j}(\eta, k)$ are four functions to be determined and ${\hat{O}}_{l,j} (k)$ are operators depending only on momentum $k$. 
As ${\hat{\chi}}_z(\vec k , \eta)$ must satisfy the Dirac equation, Eq.~\eqref{eq-momentum-rep-z}, the following constraints arise:
\begin{subequations}
	\begin{align}
	\hat O_{-1,j} &= \frac{-1}{k r_{-1,j}} \left( i r'_{1,j} - m a r_{1,j}\right) \hat O_{1,j},\label{eq:m1l}\\
	\hat O_{1,j} &= \frac{-1}{k r_{1,j}} \left( i r'_{-1,j} + m a r_{-1,j}\right) \hat O_{-1,j}.\label{eq:1l}
	\end{align}
\end{subequations}
In order to obtain these constraints, one has to multiply~Eq.~\eqref{eq-momentum-rep-z} by $\gamma^5$ and $\gamma^0$ and use the fact that the $S_{l,j}$ are orthonormal eigenvectors of $\Sigma_z$ and $\gamma^0$.

The operators $\hat O_{l,j}$ are time independent, so the $r_{l,j}$ parts in the Eqs.~\eqref{eq:m1l} and~\eqref{eq:1l} are constants in time that can be absorbed in the definition of the $\hat O_{l,j}$ operators. Hence, without any loss of generality,
\begin{equation}\label{eq:o_equal}
\hat O_{1,j} = \hat O_{-1,j} .
\end{equation} 
and 
\begin{subequations}\label{eq:r_equations}
	\begin{equation}
	r_{-1,j} = \frac{-1}{k } \left( i r'_{1,j} - m a r_{1,j}\right),
	\end{equation}
	\begin{equation}
	r_{1,j} = \frac{-1}{k } \left( i r'_{-1,j} + m a r_{-1,j}\right).
	\end{equation}
\end{subequations}
Equations~\eqref{eq:r_equations} imply that 
\begin{equation}\label{eq:r_second_equation}
r''_{\pm1,j}+\left(k^2+m^2 a^2\pm i m a'\right)r_{\pm1,j}=0.
\end{equation}

Equation~\eqref{eq:r_second_equation} admits two independent solutions. One is related to particles ($u_\pm$) and the other is related to anti-particles ($v_\pm$)\footnote{As these equations do not depend on helicity, the independent solutions of the Eqs~\eqref{eq:r_second_equation} are not expressed in terms of the helicity index $j$.}
\begin{subequations}\label{eq:part_anti}
	\begin{equation}
	r_{1,j}(\eta,k) \hat O_{1,j}(k) =  u_+(\eta,k) \hat{a}(k) +  v_+(\eta,k) \hat{b}^\dagger(-k) ,
	\end{equation}
    \begin{align}
	r_{-1,j}(\eta,k) \hat{O}_{-1,j}(k) &= r_{-1,j}(\eta,k) \hat{O}_{1,j}(k) \nonumber \\
	&=  u_-(\eta,k) \hat{a}(k) +  v_-(\eta,k) \hat{b}^\dagger(-k) ,
	\end{align}
\end{subequations}
where the $-k$ in the argument of $\hat{b}$ has its origin from the single complex exponential appearing in the definition~\eqref{momentum-rep}.

With the definitions of $u_\pm$ and $v_\pm$ in Eqs.~\eqref{eq:part_anti}, Eqs.~\eqref{eq:r_equations} read
\begin{subequations}\label{eq:u_equations}
	\begin{equation}
	u_+(k,\eta)= \frac{-1}{k}\left(i u_-'(k,\eta)-m a(\eta) u_-(k,\eta)\right),
	\end{equation}
	\begin{equation}
	u_-(k,\eta)= \frac{-1}{k}\left(i u_+'(k,\eta)+m a(\eta) u_+(k,\eta)\right),
	\end{equation}
\end{subequations}
and the same relation are valid for $v_\pm$. Due to Eqs.~\eqref{eq:u_equations}, the quantity $|u_+|^2+|u_-|^2$ is conserved in time, hence it can be normalized
\begin{equation}
|u_+|^2+|u_-|^2 = 1 .
\end{equation}

Calculating the second order derivative of Eqs.~(\ref{eq:u_equations}) and using both to decouple $u_\pm(k,\eta)$, one obtains the following second order equations analogous to Eqs.~(\ref{eq:r_second_equation}):
\begin{equation}\label{eq:upm_second_equation}
u_{\pm}''(k,\eta)+\left(k^2+m^2 a^2\pm i m a'\right)u_{\pm}(k,\eta)=0
\end{equation}

The definitions of particle and anti-particle are given in the initial conditions for $u\pm$ and $v_\pm$ at $\eta=-\infty$, when Eq.~\eqref{eq:r_second_equation} can be separated into positive (particles) and negative (anti-particles) frequencies without ambiguity.

If $u_+(k,\eta)$ is a solution of the equation~\eqref{eq:r_second_equation}, then the function $u_-^*(k,\eta)$ is a linear independent solution of the same equation\footnote{The Wronskian of $u_+$ and $u_-^*$ is constant. In fact, from \eqref{eq:u_equations}
	$$
	W(u_+,u_-^*)={u'}_+ u_-^*-u_+ {u'}_-^* = i k (|u_+|^2+|u_-|^2) = ik.
	$$}. 
It implies that, with a choice of phase,
\begin{equation}
v_\pm = \mp u^*_\mp .
\end{equation}
Hence, the functions $u_+$ and $v_+$ ($u_-$ and $v_-$) are linear independent solutions of Eq.~\eqref{eq:r_second_equation}, which represent particles and anti-particles, respectively.

With $u_\pm$ and $v_\pm$ inserted in equation Eq.~\eqref{expansion-basis}, the $z$-direction spinor operator can be separated into particle and anti-particle contributions for the two helicities values,
\begin{equation}\label{eq:xz_particle}
\hat{\chi}_z(\eta,k) = \sum_j \left(U_j \hat{a}_j(k)+V_j \hat{b}^\dagger_j(-k)\right) ,
\end{equation}
where $U_j$ ($V_j$) corresponds to the particle (anti-particle) spinor with helicity $j$
\begin{subequations}
	\begin{equation}
	U_j = u_+ S_{1,j} + u_- S_{-1,j},
	\end{equation}
	\begin{equation}
	V_j = v_+ S_{1,j} + v_- S_{-1,j}.
	\end{equation}
\end{subequations}
It can be verified that these Dirac spinors satisfy the following relation:
\begin{equation}
\label{relation-Ul-Vl}
V_j (k,\eta) = C\gamma^0\Pi^*U_j^*(k,\eta) = C {\bar{U}}^{\rm T}_j(-k,\eta) ,
\end{equation}
where $C=i\gamma^2\gamma^0$ is the charge conjugation matrix and $\Pi = -i\gamma^0\gamma^1\gamma^5$ is the parity operator in $k$ space (e.g., $\Pi U_j(k) = U_j(-k)$, where $U_j(-k)$ satisfies the Dirac equation~\eqref{eq-momentum-rep-z} with the replacement $k \rightarrow -k$). 
Equation~\eqref{relation-Ul-Vl} also appears in other references~\cite{Riotto}. 

Note also that $U_l, V_l$ satisfy the following important properties:
\begin{align*}
&U_j^\dagger V_m=0, &\quad& {\bar{U}}_j \frac{\partial}{\partial \eta} V_j - \biggl{(}\frac{\partial}{\partial \eta} {\bar{U}}_j\biggr{)}{V}_j =0, \\
&\frac{\partial}{\partial \eta}(U_j^\dagger V_m)=0, &\quad& U_j^\dagger U_m=V_j^\dagger V_m=\delta_{jm}.
\end{align*}

The Hamiltonian of the fermionic particles is given by
\begin{equation}\label{eq:hamiltonian_formal}
H = \int d^3\vec{x} \hat{\chi}^\dagger (x) \left(- i \partial_\eta\right) \hat{\chi}(x).
\end{equation}
From the canonical anti-commutation relations and the orthonormality of $U_l, V_m$, we obtain
\begin{equation}\label{anticommutation}
\{\hat{a}_j(\vec{k}),\hat{a}_m^{\dagger}(\vec{k}')\} = \{\hat{b}_j(\vec{k}),\hat{b}_m^{\dagger}(\vec{k}')\} = \delta_{jm} \delta^3(\vec{k}-\vec{k}') ,
\end{equation}
and null for the other combinations. 

Substituting Eqs.~\eqref{momentum-rep} and~\eqref{eq:xz_particle} into the Hamiltonian, Eq.~\eqref{eq:hamiltonian_formal}, we obtain:
\begin{align}\label{eq:hamiltonian_expand}
&H = \int d^3 \vec{k} \sum_{j} \left\{ E_k(\eta) \left[ \hat{a}^\dagger_j(k) \hat{a}_j(k)-\hat{b}_j(-k)\hat{b}^\dagger_j(-k)\right] + \right. \nonumber\\ 
& \left. + F_k(\eta) \hat{b}_j(-k) \hat{a}_j(k) + F^*_k(\eta) \hat{a}^\dagger_j(k) \hat{b}^\dagger_j(-k)\right\} ,
\end{align}

where
\begin{subequations}\label{E-F}
\begin{align}
&\omega_k(\eta) = \sqrt{k^2+m^2 a^2(\eta)} ,\label{wk}\\
&E_k(\eta) = k Re\left(u_+^* u_-\right)+ m a(\eta) \left(1-|u_+|^2\right) ,\label{Ek}\\
&F_k(\eta) = \frac{k}{2}\left(u_+^2- u_-^2\right) + m a(\eta) u_+ u_- ,\label{Fk}\\
&E_k^2 + |F_k|^2 = \omega_k^2 ,\qquad  -\omega_k \leq E_k \leq \omega_k \label{EkFk}.
\end{align}
\end{subequations}
One can diagonalize the Hamiltonian~\eqref{eq:hamiltonian_expand} through the Bogoliubov transformation~\cite{Parker:1971pt}:
\begin{subequations}\label{eq:bogu}
	\begin{align}
	\hat{\tilde a}_j(k,\eta) &= \alpha_k(\eta) \hat{a}_j(k) + \beta_k(\eta) \hat{b}^\dagger_l (-k) ,\\
	\hat{\tilde b}_j(k,\eta) &= -\beta^*_k(\eta) \hat{a}_j(k) + \alpha^*_k(\eta) \hat{b}^\dagger_l (-k) ,
	\end{align}
\end{subequations}
where $\alpha_k(\eta)$ and $\beta_k(\eta)$ satisfy
\begin{subequations}\label{eq:alpha_beta}
    \begin{align}
    \alpha_k(\eta)& = \beta_k (\eta)\left( \frac{E_k(\eta) + \omega_k
(\eta)}{F^*_k(\eta)}\right), \\
    \beta_k(\eta)&= \frac{F^*_k(\eta)}{2\omega_k (\eta)\alpha_k^* (\eta)},\\
    |\beta_k(\eta)|^2  &= \frac{|F_k(\eta)|^2}{2\omega_k
(\eta)(\omega_k (\eta)+E_k(\eta))}\notag\\
    &=\frac{\omega_k (\eta) - E_k(\eta)}{2 \omega_k (\eta)}  .
    \end{align}
and
\begin{align}
|\alpha_k(\eta)|^2  + |\beta_k(\eta)|^2&= 1, \\
|\alpha_k(\eta)|^2 -|\beta_k(\eta)|^2&=\frac{E_k(\eta)}{\omega_k (\eta)} .
\end{align}
\end{subequations}

In terms of the new creation and annihilation operators, Eqs.~\eqref{eq:bogu},
the normal ordered Hamiltonian operator then reads
\begin{equation}\label{hamiltonian-diagonal}
H = \int d^3 \vec{k} \sum_{j} \omega_k(\eta) \left[ \hat{\tilde a}^\dagger_j(k,\eta) \hat{\tilde a}_j(k,\eta)+\hat{\tilde b}^\dagger_j(k,\eta)\hat{\tilde b}_j(k,\eta)\right] .
\end{equation}
From the Hamiltonian~\eqref{hamiltonian-diagonal}, an observer will naturally define the vacuum state in some conformal time $\eta$ as $\hat{\tilde a}_j(k,\eta)\ket{0_{\eta}} = \hat{\tilde b}_j(k,\eta) \ket{0_{\eta}}=0$. In order to obtain the number of particles created, it is necessary to compare the different vacua in different times. This evolution is dictated by the dynamics of $u_+(\eta)$ and $u_-(\eta)$ through Eqs.~\eqref{eq:u_equations}. 
These differential equations can be rewritten in a much more clear and physical form. 
First, let us write them in a more compact way:
\begin{eqnarray}\label{matrix_m}
\mqty( u'_+ \\ u'_- )  = \mqty( -ima(\eta) & ik \\ ik & ima(\eta) )
\mqty( u_+ \\ u_- ) \leftrightarrow \mathbf{u'} = \mathbf{M}
\mathbf{u}.\quad
\end{eqnarray}

To solve this equation, we must first diagonalize the $\mathbf{M}$ matrix. 
Their orthonormal eigenvectors, with eigenvalues $\pm i\omega$, read
\begin{equation}\label{eigenvectors}
\mathbf{e_1} \equiv \begin{pmatrix} \sqrt{\frac{1}{2}(1-\frac{ma}{w})} \\ \sqrt{\frac{1}{2}(1+\frac{ma}{w})} \end{pmatrix} ,\quad \mathbf{e_2} \equiv \begin{pmatrix} \sqrt{\frac{1}{2}(1+\frac{ma}{w})} \\ - \sqrt{\frac{1}{2}(1+\frac{ma}{w})} \end{pmatrix}.
\end{equation}
Defining the vector
\begin{widetext}
\begin{equation}\label{vector-z}
{\mathbf{z}} \equiv \begin{pmatrix} \alpha_k \\ \beta_k \end{pmatrix} \equiv \begin{pmatrix} e^{-i\int \omega d\eta} & 0 \\ 0 & e^{i\int \omega d\eta} \end{pmatrix} \begin{pmatrix} \sqrt{\frac{1}{2}(1-\frac{ma}{w})} & \sqrt{\frac{1}{2}(1+\frac{ma}{w})} \\ \sqrt{\frac{1}{2}(1+\frac{ma}{w})} & -\sqrt{\frac{1}{2}(1-\frac{ma}{w})} \end{pmatrix}\begin{pmatrix} u_+ \\ u_- \end{pmatrix},
\end{equation}
\end{widetext}
we obtain the following differential equation for $\mathbf{z}$:
\begin{equation}\label{z-equation}
{\mathbf{z}}' = \begin{pmatrix} {\alpha_k}' \\ {\beta_k}' \end{pmatrix} = \frac{ma'k}{\omega^2}\begin{pmatrix} 0 & - e^{-2i\int \omega d\eta} \\ e^{2 i\int \omega d\eta}& 0 \end{pmatrix} \begin{pmatrix} \alpha_k\\ \beta_k\end{pmatrix}  . 
\end{equation}
The functions $\alpha_k (\eta)$ and $\beta_k (\eta)$ are exactly the same as the ones defined in Eq.~\eqref{eq:bogu}. In order to see this, we must invert ${\mathbf{z}}$ using Eq.~\eqref{vector-z} to obtain
\begin{subequations}\label{u+u-}
	\begin{align}
	u_{k,+} (\eta)&= \alpha_k (\eta)\sqrt{\frac{1}{2}\left(1-\frac{ma}{\omega}\right)} \exp\qty{i\int \omega d\eta} + \nonumber \\
	&+ \beta_k (\eta) \sqrt{\frac{1}{2}\left(1+\frac{ma}{\omega}\right)} \exp\qty{-i\int \omega d\eta} ,\\
	u_{k,-} (\eta)&= \alpha_k (\eta)\sqrt{\frac{1}{2}\left(1+\frac{ma}{\omega}\right)} \exp\qty{i\int \omega d\eta} + \nonumber \\
	&-\beta_k (\eta)\sqrt{\frac{1}{2}\left(1-\frac{ma}{\omega}\right)} \exp\qty{-i\int \omega d\eta} ,
	\end{align}
\end{subequations}
and substitute them in Eqs.~\eqref{Ek} and \eqref{Fk} to get Eqs.~\eqref{eq:alpha_beta}\footnote{Due to our signature convention, positive frequencies are associated with particles.}.

From Eq.~\eqref{z-equation}, one can see that when $ma'k/\omega^2=mHk/(m^2 + k^2/a^2)$ becomes negligible, $\alpha_k (\eta)$ and $\beta_k (\eta)$ tends to be constant. This happens in a flat universe, or whenever $H$ becomes negligible in some FLRW model. 
This is the case of bouncing models in which the far past of the contracting phase is driven by a fluid satisfying the strong energy condition. 
In this era, the above quantities become constant, and if we choose $\beta_k = 0$, which implies $F_k=0$ and $E_k=\omega$, the Hamiltonian~ \eqref{eq:hamiltonian_expand} assumes the form of Hamiltonian~\eqref{hamiltonian-diagonal}, and we have a stable vacuum defined by 
${\hat{a}}_l(k)\ket{0} = {\hat{b}}_l(k)\ket{0}=0$. In this situation, when $\alpha_k=1$ and $\beta_k=0$, we can set the initial conditions for the modes $u_\pm (\eta)$ using Eqs.~\eqref{u+u-}, yielding
\begin{equation}\label{ICS_u+u-}
u_{k,\pm}(\eta_0)= \sqrt{\frac{1}{2}\left(1\mp\frac{ma(\eta_0)}{\omega (\eta_0) }\right)} e^{i\phi},
\end{equation}
where $\phi$ is an arbitrary phase.

In the expanding phase, where the observer defines the new vacua 
$\ket{0_{\eta}}$, the particle number operator 
$n_l(k)={\hat{a}}_l^\dagger(k){\hat{a}}_l(k)$
will give the average number of particles with momentum $k$ in the vacuum state $\ket{0_{\eta}}$:
\begin{equation}
\langle n_k (\eta)\rangle_0=\sum_l\bra{0_{\eta}}n_l(k)\ket{0_{\eta}}=2|\beta_k(\eta)|^2.
\end{equation}
The total particle number density $n(\eta)$ can be defined as the limit of $\sum_{k}|\beta_k(\eta)|^2$ in a box of side $L\to\infty$ divided by the volume $V(\eta)=(a(\eta)L)^3$, which reads
\begin{align}\label{n}
n(\eta)&=\left(\frac{1}{a(\eta)^3L^3}\right)\left(\frac{L}{2\pi}\right)^3\int\limits_0^\infty d^3k  \langle n_k (\eta)\rangle_0 \nonumber \\
&=\frac{1}{\pi^2a(\eta)^3}\int\limits_0^\infty dk \ k^2 |\beta_k(\eta)|^2,
\end{align}
and the same for the anti-particles. 
In addition, we can define the energy density of created particles from an analogous momentum sum of $\langle n_k (\eta)\rangle_0\ \omega_k(\eta)$, which results
\begin{align}\label{rho}
\Delta\rho(\eta)=\frac{1}{\pi^2a^4(\eta)}\int\limits_0^\infty dk\  k^2 |\beta_k(\eta)|^2\omega_k(\eta),
\end{align} 
where $\omega_k(\eta)$ is given by Eq.~\eqref{wk} and $\Delta\rho (\eta)$ is the energy density of created particles at the instant of time $\eta$.

Concluding this section, our task in the following will be to solve Eq.~\eqref{z-equation} with initial condition 
\begin{equation}\label{initial-condition}
{\mathbf{z}}(\eta=\eta_i) = \begin{pmatrix} 1 \\ 0 \end{pmatrix} \quad , 
\end{equation}
where $\eta_i$ is some initial conformal time in the far past of the bouncing model of interest, and find $|\beta_k^2(\eta)|$ in order to obtain $n(\eta)$ and $\Delta\rho(\eta)$.

\section{The Background Bouncing Model}
\label{sec3}

We consider fermion production in bouncing cosmological models described by quantum cosmology in the light of the de Broglie-Bohm interpretation of quantum mechanics. 
In these models, quantum effects are responsible for the avoidance of the classical cosmological singularity by the presence of quantum corrections to the classical Friedmann equations.
The background is quantized using the Wheeler-deWitt framework, where the phase of the wave-function guides the background evolution (see Ref.~\cite{Pinto-Neto:2013toa} and references therein).

Quantum fields in an expanding background have an ambiguous definition for its vacuum states~\cite{Kinney:2005in}. This ambiguity is due to the lack of a general procedure to define an unique set of Fourier modes when the space-time does not have a global time-like killing vector. 
However, it is possible to choose a suitable vacuum state, called {\it adiabatic vacuum}~\cite{BirrelDaviesBook,ParkerTomsBook}, for which the expectation value of the number operator varies slowly as the expansion rate of the Universe is arbitrarily slow.

In an expanding background, for each instant of time we can define a complete set of solutions for the Fourier modes, which defines creation and annihilation operators and, consequently, a vacuum state~\cite{BirrelDaviesBook,WaldBook,Chung:2003wn}. 
For this reason, if one compares two different vacua at, e.g., instants $t_i$ and $t_f$, evolving the vacuum defined at $t_i$ until $t_f$ will not correspond to the same vacuum originally defined at $t_f$. 
Actually, the creation and annihilation operators at different times are related by a Bogoliubov transformation, which indicates creation of particles. Thus, particles can be created in the presence of a gravitational field~\cite{ParkerTomsBook}, as it was explicitly shown in the previous section for fermions. 

In a Friedmann universe with a bounce solution, the presence of bounce physics implies a large deviation from Minkowski space-time due to its strong gravitational field. 
Thus, for vacua defined at $t_i$ and $t_f$, $t_f>t_i$, the expectation value at $t_f$ of the particle number operator defined at $t_i$ may result in substantial particle creation in the contracting phase and through the bounce, which may become relevant to the dynamical evolution of the background model and to baryogenesis. 
Scalar particle production in this context has already been explored in Ref.~\cite{Celani:2016cwm} for a variety of regimes, showing that particle creation cannot be neglected in some cases. 
We will give a step forward and calculate the production of fermions in this same context.
Scalar particle production has also been explored in other bounce models, like matter-bounce cosmology~\cite{Quintin:2014oea}, loop quantum cosmology~\cite{Haro:2015zda,Tavakoli:2014mra} and in the new Ekpyrotic model~\cite{Hipolito-Ricaldi:2016kqq}, whereas fermion production has also been investigated in the context of superstring cosmology~\cite{Tsujikawa:2001ud}.

In the next subsection we contextualize fermion production in cosmological inflation, which is an alternative to the bounce cosmology scenario, or even coexistent. In the following, we introduce fermion production in bouncing models.

\subsection{Inflation}
The inflationary scenario, in its most common implementation, is realized by a scalar field slowly rolling down its potential~\cite{Lyth:1998xn}. By the time the inflationary quasi-de Sitter phase comes to an end, the universe is still unpopulated by particles. The mechanism responsible for the particle production in the scenario is the aforementioned parametric amplification, during the so-called preheating and reheating phases~\cite{Linde:94,Linde:97,Amin_Reheating}. 

The reheating consists in the decay of the inflaton field through oscillations around its minimum. The coupling of the inflaton to bosonic and/or fermionic fields allows its decay to the respective bosons and/or fermions. Each reheating model has its peculiarities~\cite{Amin_Reheating,Brandenberger_Reheating}, but they must not contradict the predictions of Big-Bang Nucleosynthesis.

Particle production can be even more efficient considering a phase prior to the reheating\footnote{It can also be thought as the first phase of reheating.}. Contrary to the narrow parametric resonance of the reheating, a broad resonance can be achieved in considering non-perturbative effects on the inflaton field. The preheating phase~\cite{Linde:94,Brandenberger_Reheating} then opens new channels of decay, boosting the production of particles.

Relevant to this work, the fermion production during preheating is developed in the aforementioned papers~\cite{Riotto,Peloso:2000hy}. The focus on high mass fermions was given to their ensuing leptogenesis modeling, which characterizes those heavy fermions as Majarona right-handed neutrinos. The decay of such particles generates the desired B - L non-conservation, necessary for the baryon asymmetry from leptogenesis. These kind of neutrinos can also take part in the see-saw mechanism~\cite{GellMann:1980vs,Neutrino_mass}, responsible for the low mass of SM neutrinos. Similar use of right-handed neutrinos is encountered throughout the literature, in different kinds of models~\cite{Ballesteros:2016xej}.

\subsection{Bounce}

The Wheeler-DeWitt equation for a minisuperspace model of a FLRW geometry in the case where the matter content is a single hydrodynamical fluid with a barotropic equation $p=\lambda\rho$ is given by
\begin{align}\label{WDW}
i\frac{\partial\Psi_{(0)}(a,T)}{\partial T}
=
\frac{1}{4}\frac{\partial^2\Psi_{(0)}(a,T)}{\partial \chi^2},
\end{align} 
where
\begin{align}
\chi=\frac{2}{3}(1-\lambda)^{-1}a^{3(1-\lambda)/2},
\end{align} 
$a$ is the scale factor and $T$ is a degree of freedom which plays the role of time.
The solution for this equation~\cite{AcaciodeBarros:1997gy,Alvarenga:2001nm} is 
\begin{widetext}
\begin{align}\label{psi}
&\Psi_{(0)}(a,T)=\left[\frac{8T_b}{\pi\left(T^2+T_b^2\right)}\right]^{1/4}\exp{\left[\frac{-4T_ba^{3(1-\lambda)}}{9\left(T^2+T_b^2\right)(1-\lambda)^2}\right]} \nonumber \\
&\times\exp{-i\left[\frac{4Ta^{3(1-\lambda)}}{9\left(T^2+T_b^2\right)(1-\lambda)^2}+\frac{1}{2}\arctan{\left(\frac{T_b}{T}\right)}-\frac{\pi}{4}\right]}.
\end{align} 
\end{widetext}
It satisfies an unitary evolution condition and it comes from the normalized initial wave-function
\begin{align}\label{psi_0}
\Psi_{(0)}^{(i)}(\chi)=\left(\frac{8}{T_b\pi}\right)^{1/4}\exp{-\frac{\chi^2}{T_b}}.
\end{align} 
The probability density $\rho(a,T)=a^{(1-3\lambda)/2}\left|\Psi_{(0)}(a,T)\right|^2 $ satisfies a continuity equation
\begin{align}\label{prob_density}
\frac{\partial\rho}{\partial T}-\frac{\partial}{\partial a}\left[\frac{a^{(3\lambda-2)}}{2}\frac{\partial S}{\partial a}\rho\right]=0.
\end{align}
Recalling the de Broglie-Bohm quantum theory~\cite{HollandBook}, the usual Schr\"odinger equation for a non-relativistic particle in the coordinate representation reads
\begin{align}\label{schrodinger}
\frac{\partial \Psi({\bf x},t)}{\partial t}=\left[-\frac{\hbar^2}{2m}\nabla^2+V({\bf x})\right]\Psi({\bf x},t).
\end{align}
Expressing $\Psi=R e^{iS/\hbar}$ and substituting in Eq.~(\ref{schrodinger}), two equations are obtained. The evolution of the probability density $R^2$ is given by the continuity equation
\begin{align}\label{schrodinger_prob}
\frac{\partial R^2}{\partial t} + \nabla\cdot\left(R^2 \frac{\nabla S}{m}\right)=0,
\end{align}
where one identifies $v={\nabla S}/m$ as the velocity field of the position of the particle, which is assumed to have objective reality.
Comparing Eqs.~(\ref{prob_density}) and~(\ref{schrodinger_prob}), and assuming that in General Relativity (GR) it is the metric amplitude which is assumed to have objective reality (in this simple case it is just the scale factor), one obtains the evolution equation for the scale factor, given by
\begin{align}\label{da_dT}
\frac{da}{dT}=-\frac{a^{(3\lambda-2)}}{2}\frac{\partial S}{\partial a}.
\end{align}
Calculating $\partial S/\partial a$ from Eq.~(\ref{psi}), the solution for $a(T)$ reads
\begin{align}\label{a_T}
a(T)=a_b\left[1+\left(\frac{T}{T_b}\right)^2\right]^{1/[3(1-\lambda)]},
\end{align}
which is nonsingular at $T=0$ and tends to the classical solution for $T\to\pm\infty$.

We are interested in the more usual fluids, which are radiation and dust matter.
From our knowledge of classical Friedmann cosmology, radiation dominated for small $a$, so it will dominate during the bounce. Dust matter dominated far from the bounce, then we choose to consider in this work the cases for pure radiation and radiation plus dust matter.
We are using the time gauge $N=a^{3\lambda}$, for which $NdT=ad\eta$, where $\eta$ is the conformal time. Therefore, $\eta$ is given in terms of $T$ as
\begin{align}
d\eta=\left[a(T)\right]^{3\lambda-1}dT.
\end{align}
For pure radiation ($\lambda=1/3$), we obtain $T=\eta$ and by Eq.~(\ref{a_T}) for this particular case the scale factor reads
\begin{align}\label{a_eta_rad}
a(T)=a_b\sqrt{1+\left(\frac{\eta}{\eta_b}\right)^2}.
\end{align}
For radiation plus dust matter, the calculation is given in detail in Ref.~\cite{PintoNeto:2005gx} and the scale factor is given by
\begin{align}\label{a_eta_rad_dust}
a(\eta)=a_e\left[\left(\frac{\eta}{\eta_*}\right)^2+2\frac{\eta_b}{\eta_*}\sqrt{1+\left(\frac{\eta}{\eta_b}\right)^2}\right],
\end{align} 
where $a_e$ is the scale factor at matter-radiation equality, and the parameters $\eta_*$ and $\eta_b$ are related to the wave-function parameters. We recover the case of pure radiation expanding this expression for large $\eta_*$ and identifying $a_b=2a_e\eta_b/\eta_*$.

In order to make contact with cosmological data, it is convenient to reparametrize the bounce solutions in terms of observable quantities. 
The Friedmann equation for radiation and dust matter fluids reads
\begin{align}\label{Hubble_Omega}
H^2=H_0^2\left(\frac{\Omega_{r0}}{a^{4}}+\frac{\Omega_{m0}}{a^{3}}\right),
\end{align} 
where $H$ is the Hubble parameter, $\Omega_r=\rho_r/\rho_{\mathrm{crit}}$ and $\Omega_m=\rho_m/\rho_{\mathrm{crit}}$ are the density parameters for radiation and dust matter, respectively,
and $\rho_{\mathrm{crit}}=3H^2/(8\pi G)$ is the critical density.
The subscript '$0$' denotes the values we observe today.
The critical density today is $\rho_{\mathrm{crit0}}\approx   10^{-29}\ \mathrm{g/cm^3}$.

Far from the bounce scale (large $\eta$), where quantum effects are negligible, the Friedmann equation reads
\begin{align}\label{Hubble_a}
H^2=\frac{4a_e}{\eta_*^2}\left(\frac{a_e}{a^{4}}+\frac{1}{a^{3}}\right),
\end{align} 
Comparing Eqs.~(\ref{Hubble_Omega}) and~(\ref{Hubble_a}), in terms of the comoving Hubble radius $R_H=1/(a_0H_0)$, the density parameters today are given by
\begin{align}\label{Omegas}
\Omega_{r0}=\frac{a_e}{a_0}\frac{4R_H^2}{\eta_*^2}, \ \ \ \ \Omega_{m0}=\left(\frac{a_e}{a_0}\right)^2\frac{4R_H^2}{\eta_*^2}.
\end{align} 
Expanding the scale factor~(\ref{a_eta_rad_dust}) for large $\eta_*$, {\it i.e.}, for radiation domination near the bounce and dust matter domination in the far past, the Friedmann equation results
\begin{align}\label{Omegas}
H^2=H_0^2\Omega_{r0}x^4\left(1-\frac{x^2}{x_b^2}\right),
\end{align} 
where $x=a_0/a$ is a redshift variable and, consequently, $x_b$ provides the redshift where the bounce occurs (apart from a small correction from dust matter density), which is defined by
\begin{align}\label{xb}
x_b=\frac{R_H}{\eta_b\sqrt{\Omega_{r0}}}.
\end{align} 
Solving Eqs.~(\ref{Omegas}) and ({\ref{xb}}) for $a_e$, $\eta_*$ and $\eta_b$, and computing the scale factor at the bounce $a_b$ in terms of theses quantities, one obtains
\begin{align}\label{ae_eta*_etab}
&a_e=a_0\frac{\Omega_{r0}}{\Omega_{m0}},\ \ \ \ \eta_*=2R_H\frac{\sqrt{\Omega_{r0}}}{\Omega_{m0}}, \nonumber \\
&\eta_b=\frac{R_H}{x_b\sqrt{\Omega_{r0}}}, \ \ \ \ a_b=\frac{a_0}{x_b}.
\end{align} 
In terms of these variables, the bounce curvature scale can be obtained from the four-dimensional Ricci scalar $R=6a''(\eta)/a^3(\eta)$, which results in
\begin{align}\label{Lb}
L_b&=\left.\frac{1}{\sqrt{R}}\right|_{\eta=0}=\frac{a_b\eta_b}{\sqrt{6(1+2\gamma_b)}} \nonumber \\
&=\frac{1}{\sqrt{1+2\gamma_b}}\frac{a_0R_H}{x_b^2\sqrt{6\Omega_{r0}}},
\end{align} 
where 
\begin{align}\label{gamma_b}
\gamma_b\equiv\frac{\Omega_{m0}}{4x_b\Omega_{r0}}.
\end{align} 
is the ratio of the dust matter and radiation density at the bounce.
The bounce depth value $x_b$, which appears explicitly in the Friedmann equation, 
must be constrained by physical conditions. 
The first condition is that the bounce curvature scale must be larger that the Planck length, $L_b>L_p$, which sets an upper bound on $x_b$. 
This bound is relevant since the Wheeler-DeWitt equation should be a valid approximation for any fundamental quantum gravity theory only at scales not so close to the Planck length.
Using $H_0=70\ [\text{Km}\ \text{s}^{-1}\ \text{Mpc}^{-1}]$, we obtain $a_0R_H/L_p\approx 8\times 10^{60}$, which sets
\begin{align}\label{xb_upper}
x_b\lesssim \frac{\sqrt{8}10^{30}}{(6\Omega_{r0})^{1/4}}\approx 2\times 10^{31}.
\end{align} 
This result is obtained for $\gamma_b\ll1$, where one assumes the bounce energy scale must be larger than at the start of nucleosynthesis ($\approx 10$ \text{MeV}). 
We have assumed that $\Omega_{r0}$ should not be smaller than its usual value $\Omega_{r0}=8\times10^{-5}$, and we used the cosmic microwave background radiation temperature value $T=2.7\ K$.
This assumption on the energy scale yields a second condition $x_b\gg 10^{11}$, which gives an lower bound in the bounce depth.
Therefore, we obtain the constraint
\begin{align}\label{xb_range}
10^{11}\ll x_b\lesssim 2\times 10^{31}.
\end{align} 
In the case where dust matter is taken into account, assuming the value $\Omega_{m0}\approx 0.3$, from the range of $x_b$ one obtains the following interval for $\gamma_b$: 
\begin{align}\label{gammab_range}
3.7\times10^{-29}\lesssim \gamma_b\ll 7.5\times 10^{-9}.
\end{align} 
The small values of $\gamma_b$ make it explicit that the dust matter fluid dominates only in the far past, whereas the radiation fluid dominates near the bounce scale. 

Some of the bounce parameters introduced above appear explicitly in the equations of motion of fermions in the Friedmann background with bouncing. For this reason, it is convenient to introduce some new parameters in terms of the current ones to be used in these equations in the following sections, which are defined by
\begin{align}\label{parameters}
\bar{\eta}=\frac{\eta}{\eta_b},\ \ \ \ \ \bar{k}=k\eta_b,\ \ \ \ \ r_b=ma_b\eta_b.
\end{align} 
In terms of these parameters, the scale factor, Eq.~({\ref{a_eta_rad_dust}}), for radiation and dust matter can be written as
\begin{align}\label{a_etabar_rad_dust}
a(\bar{\eta})=a_b\left(\gamma_b\bar{\eta}^2+\sqrt{1+\bar{\eta}^2}\right),
\end{align} 
whereas in the case of pure radiation ($\Omega_{m0}=0$, {\it i.e.}, $\gamma_b=0$) it reduces to
\begin{align}\label{a_etabar_rad}
a(\bar{\eta})=a_b\sqrt{1+\bar{\eta}^2}.
\end{align} 
Finally, it is relevant to notice from Eq.~(\ref{Lb}) that $L_b\approx a_b\eta_b$, hence  
\begin{align}\label{rb}
r_b\approx\frac{L_b}{L_C},
\end{align} 
where $L_C\equiv 1/m$ can be identified with the Compton length of the massive particle.

In the following subsection, we will introduce the equations of motion for fermions in the background bouncing models presented above.

\subsection{Equations}

The equations of motion (\ref{eq:upm_second_equation}) for the variables $u_{k,\pm}(\eta)$ in the bouncing background read in terms of the parameters (\ref{parameters}) read
\begin{align}\label{eqs_u}
\frac{d^2 u_{\bar{k},\pm}(\bar{\eta})}{d\bar{\eta}^2}+\left[\bar{k}^2+\frac{r_b^2}{a_b^2}a(\bar{\eta})^2 \pm i\frac{r_b}{a_b}\frac{da(\bar{\eta})}{d\bar{\eta}}\right]u_{\bar{k},\pm}(\bar{\eta})=0,
\end{align} 
where initial conditions for $u_{\bar{k},\pm}(\eta)$, Eqs.~(\ref{ICS_u+u-}), in the new variables read
\begin{align}\label{ics_u}
u_{\bar{k},\pm}(\bar{\eta}_0)=\sqrt{\frac{1}{2}\left(1\mp\frac{r_b a(\bar{\eta}_0)}{a_b\omega(\bar{\eta}_0)}\right)}e^{i\phi}.
\end{align} 
In the special case where the universe matter content has only a radiation fluid, the scale factor is given by Eq.~(\ref{a_etabar_rad}). Hence, Eq.~(\ref{eqs_u}) reads
\begin{align}\label{eqs_u_rad}
\frac{d^2 u_{\bar{k},\pm}(\bar{\eta})}{d\bar{\eta}^2}
+
\left[
\bar{k}^2
+
r_b^2\left(1+\bar{\eta}^2\right)
\pm 
\frac{ir_b\bar{\eta}}{\sqrt{1+\bar{\eta}^2}}
\right]
u_{\bar{k},\pm}(\bar{\eta})=0.
\end{align} 
These equations have no analytical solutions in terms of known functions and need to be solved numerically. It is worth mentioning that its asymptotic limits ($\bar{\eta}\to\pm\infty$) have solutions in terms of parabolic cylinder functions~\cite{gradshteyn}. These same special functions give analytical results for the Fourier modes of a scalar field in the same background, which have similar equations except for the presence of the complex term in the square brackets.

In the case where the energy fluid content is radiation and dust matter, the scale factor is given by Eq.~(\ref{a_etabar_rad_dust}), and Eq.~(\ref{eqs_u}) results in
\begin{widetext}
\begin{align}\label{eqs_u_rad_mat}
\frac{d^2 u_{\bar{k},\pm}(\bar{\eta})}{d\bar{\eta}^2}
+
\left[
\bar{k}^2
+
r_b^2\left(\gamma_b\bar{\eta}^2 +\sqrt{1+\bar{\eta}^2}\right)^2
\pm 
ir_b\bar{\eta}\left(2\gamma_b+\frac{1}{\sqrt{1+\bar{\eta}^2}}\right)
\right]
u_{\bar{k},\pm}(\bar{\eta})=0.
\end{align}
\end{widetext}
These equations have no analytical solution as well, and are solved numerically. Asymptotically analytical solutions are also no longer available.

Once the solutions for $u_{\bar{k},\pm}(\bar{\eta})$ are obtained, the occupation number $|\beta_{\bar{k}}(\bar{\eta})|^2$ for each mode $\bar{k}$ can be obtained from Eq.~(\ref{eq:alpha_beta}).
The occupation number is a function of the rescaled conformal time $\bar{\eta}$, and we are interested in the resulting particle creation after the bounce. 
In the following section, we will see that for some momenta and masses $|\beta_{\bar{k}}(\bar{\eta})|^2$ exhibits a peak near the bounce and oscillates until stabilizing to a constant value for some $\bar{\eta}=\bar{\eta}_*$. 
From then on, particle production becomes negligible~\footnote{In Ref.~\cite{Celani:2016cwm}, one obtains analytically particle production between two asymptotic states, which are adiabatic vacua}. Therefore, for $|\beta_{\bar{k}}(\bar{\eta})|^2$ evaluated at $\bar{\eta}=\bar{\eta}_*$, one obtains the asymptotic particle number density $|\beta_{\bar{k}}|^2$, where we suppress the time variable. In terms of the parameters defined in (\ref{parameters}), Eqs.~(\ref{n}) and~(\ref{rho}) read
\begin{align}\label{n_}
n&=\frac{1}{\pi^2 a^3\eta_b^3}\int\limits_0^\infty d\bar{k} {\bar{k}}^2 |\beta_{\bar{k}}|^2,\\
\label{rho_}
\Delta\rho&=\frac{1}{\pi^2 a^4\eta_b^4}\int\limits_0^\infty d\bar{k} \bar{k}^2 |\beta_{\bar{k}}|^2\omega_{\bar{k}},
\end{align} 
where $\omega_{\bar{k}}=\sqrt{\bar{k}^2+r_b^2 a^2/a_b^2}$.

\section{Numerical Integration}
\label{sec4}

In this section we show some numerical results for the creation of neutral fermionic particles in a quantum bounce.
Information about particle creation is obtained from the behavior of the Bogoliubov coefficient $\beta_{\bar{k}}$, which is non-zero when particles are created.
From the definition of particle number density, Eq.~(\ref{rho_}), the relevant physical quantity is the integrand, from which we obtain the density of created particles for each mode $\bar{k}$. 
We performed a numerical analysis of this integrand in the logarithmic scale. 

For the fermion production during the bounce, we will focus on massive neutral particles: the Standard Model (SM) neutrinos with mass $m_{\nu}$, and the neutron, despite not being an elementary particle. In addition, we also consider heavier neutrinos present in extensions of the Standard Model motivated by~(\cite{Riotto}) in order to investigate leptogenesis.
We are particularly interested if a relevant fraction of heavier particles are possible to be created due to bounce physics.

The neutron mass is known for decades, and its value to the decimal place is $939.6 \mathrm{MeV}$. Most recent measurements of the SM neutrino masses give only upper limits to its value, of about $10^{-1} \; \mathrm{eV}$.

\subsection{Analytical considerations}

The efficiency of the production of fermions is expected to be related to their masses and to the depth of the bounce. Inspecting Eq.~\eqref{eqs_u_rad_mat}, we see that it will depend mainly on their ratio, quantified in $r_{b} \approx m L_{b} = L_{b}/L_{C}$. The parameter $\gamma_b$ is small and is effective only at matter domination, when the curvature of space-time is small and hence with negligible particle creation. Note that if $r_b=0$ Eq.~\eqref{eqs_u_rad_mat} reduces to a time-independent free harmonic oscillator equation for each mode, with no particle production. Hence, as larger is $r_b$, greater will be the production. Note that $r_b$ is usually a small number for the parameters considered here, except for very large fermion masses, see Table~1, ranging from $10^{4}$ to $10^{-28}$. 
\begin{table}[htb!]
	\renewcommand*{\arraystretch}{1.1}
	\begin{center}
		\begin{tabular}{|c|l|l|}
			\hline
			\backslashbox{$m$}{$L_b$}& $10^{-18}$ cm & $10^{-30}$ cm \\ \hline
			$10^{-10}$ GeV& $10^{-15}$ & $10^{-27}$ \\ \hline
			$1$ GeV       & $10^{-5}$ & $10^{-17}$  \\ \hline
			$10^{3}$  GeV & $10^{-2}$  & $10^{-14}$ \\ \hline
			$10^{6}$  GeV & $10^{2}$ & $10^{-11}$   \\ \hline
			$10^{9}$  GeV & $10^{4}$   & $10^{-8}$  \\ \hline
		\end{tabular}
		\caption{Order of magnitude of $r_b=mL_b$ for different masses and bounce length scale.}
	\end{center}
\end{table}

As we commented above, the presence of dust is not relevant for the production of fermions. We verified this numerically. Note that for scalar particles some differences may appear in the infrared limit, see Ref.~\cite{Celani:2016cwm}. The difference is that particle production of scalar particles depends on the second derivative of the scale factor, while fermion production depends on its first derivative. As for large scale factors $a(\eta) \approx \eta^2$ and $a(\eta) \approx \eta$ for matter domination and radiation dominated, respectively, the difference in $a''/a$ and $a'/a$ for matter domination and radiation domination have different asymptotic behaviors in the case of scalar particles and the same asymptotic behavior in the case of fermions.

In the Appendix we present estimations of the infrared and ultraviolet limits. We verified that in the infrared limit the Bogoliubov coefficient $\beta_{\bar{k}}$ goes linearly with $\bar{k}$ while in the ultraviolet limit it decreases as ${\bar{k}}^{-2}$.

\subsection{Numerical results}

\begin{figure}[htb!]
	\includegraphics[scale=0.9]{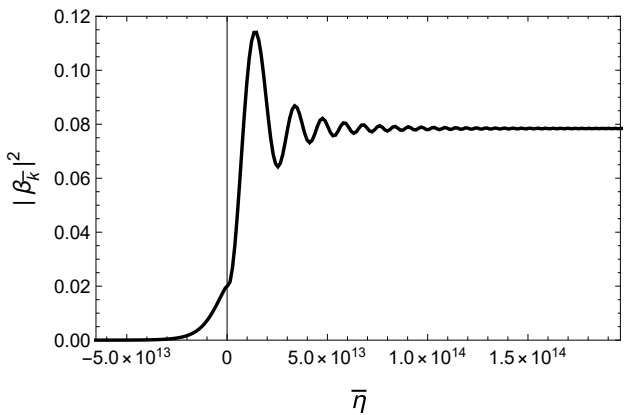}
	\caption{Plot of number density $|\beta_{\bar{k}}|^2$ for $m=10^{-1}$eV for the representative momentum  $\bar{k}=10^{-13}$ choosing $r_b=5.1\times 10^{-27}$.}  
	\label{beta2k_example}
\end{figure}

\begin{figure}[htb!]
	\includegraphics[scale=0.9]{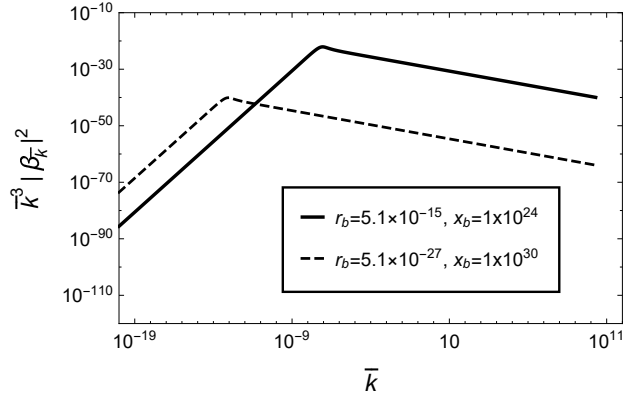}
	\caption{Logarithmic plot of the fermion number density ${\bar{k}}^3|\beta_{\bar{k}}|^2$ for the neutrino mass $10^{-1}$eV for the representative choices $x_b=10^{24}$ $\left(r_b = 5.1\times 10^{-15}\right)$ and $x_b=10^{30}$ $\left(r_b = 5.1\times 10^{-27}\right)$ given by solid and dashed lines, respectively.}  
	\label{bogoliubov_neutrinos}
\end{figure}

\begin{figure}[htb!]
	\includegraphics[scale=0.9]{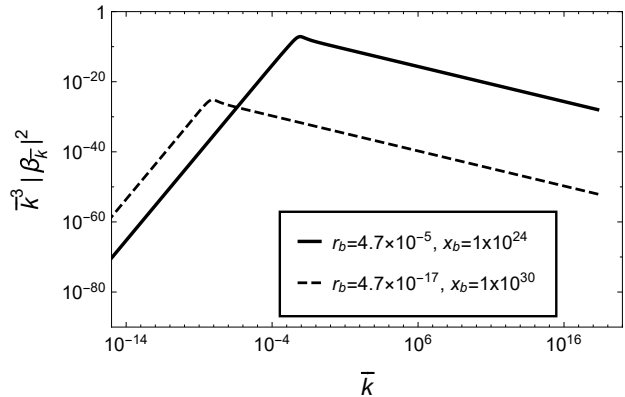}
	\caption{Logarithmic plot of the fermion number density ${\bar{k}}^3|\beta_{\bar{k}}|^2$ for the neutron mass $1$GeV for the representative choices $x_b=10^{24}$ $\left(r_b = 4.7\times 10^{-5}\right)$ and $x_b=10^{30}$ $\left(r_b = 4.7\times 10^{-17}\right)$ given by solid and dashed lines, respectively.}  
	\label{bogoliubov_neutrons}
\end{figure}

Before we present our main results, we have mentioned in the previous section that when particle production occurs the number density $|\beta_{\bar{k}}|^2$ for each mode $\bar{k}$ stabilizes to a constant value at some asymptotic instant $\bar{\eta}=\bar{\eta}_*$ after the bounce.  
In Fig.~\ref{beta2k_example} we plot $|\beta_{\bar{k}}|^2$ as a function of the conformal time $\bar{\eta}$ for the mode $\bar{k}=10^{-13}$ choosing $r_b=5.1\times 10^{-27}$ as an example that illustrates it.
In the following results, we consider $|\beta_{\bar{k}}(\bar{\eta}_*)|^2$ as the resulting particle production per mode. 
In Figs.~\ref{bogoliubov_neutrinos} and~\ref{bogoliubov_neutrons}
we plot the behavior of the logarithm of the number density of fermions in $\bar{k}$-space $\bar{k}^3|\beta_{\bar{k}}|^2$ as a function of $\ln(\bar{k})$ for the production of neutrinos and neutrons, respectively. The value of the peak of this figure yields an estimate of the integral appearing in Eq.~\eqref{n_} yielding the number density of fermions produced. 
For each case, the solid and dashed curves represent the choices $x_b=10^{24},10^{30}$, respectively. 
The neutrinos masses are not precisely known, but have the upper limit~$\leq 0.12$eV (see Ref.~\cite{Mertens:2016ihw}). 
We chose $m_\nu\approx 0.1$eV for our calculations. 
On the other hand, the neutron mass is well know, so we consider $m_n \approx 1$GeV. Note that fermion production increases with $r_b$, as we anticipated, and the infrared and ultraviolet limits are in accordance with the estimations presented in the Appendix.

\begin{figure}[htb!]
	\includegraphics[scale=0.9]{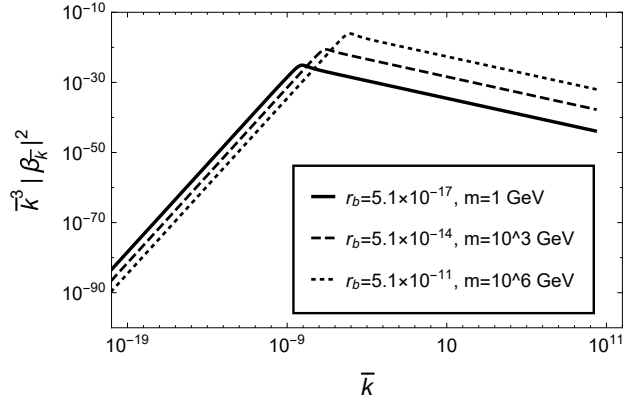}
	\caption{Logarithmic plot of the fermion number density ${\bar{k}}^3|\beta_{\bar{k}}|^2$ for the neutrino masses $1$ GeV, $10^3$ GeV, and $10^6$ GeV for the representative choice $x_b=10^{30}$ given by solid, dashed and dotted lines, respectively.}  
	\label{bogoliubov_heavy_neutrinos}
\end{figure}

\begin{figure}[htb!]
	\includegraphics[scale=0.9]{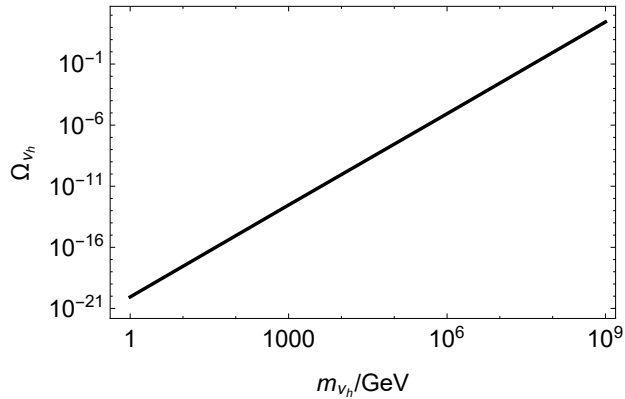}
	\caption{The density parameter $\Omega_{\nu_{\mathrm{h}}}=\rho_{\nu_{\mathrm{h}}}/\rho_{\mathrm{crit0}}$ today for heavy neutrinos normalized by $x^3$ as a function of $m_{\nu_{\mathrm{h}}}$ for the representative choice $x_b=10^{30}$.}  
	\label{OMEGA_heavy_neutrinos}
\end{figure}

In Fig.~\ref{bogoliubov_heavy_neutrinos}
we plot the behavior of the logarithm of the fermion number density in $\bar{k}$-space ${\bar{k}}^3|\beta_{\bar{k}}|^2$ as a function of $\ln(\bar{k})$ for the production of heavy neutrinos masses $1, 10^3, 10^6$GeV for the specific bounce with depth choice $x_b=10^{30}$, 
whereas in Fig.~\ref{OMEGA_heavy_neutrinos} we plot a curve fitting the density parameter $\Omega_{\nu_{\mathrm{h}}}=\rho_{\nu_{\mathrm{h}}}/\rho_{\mathrm{crit0}}$ for $x=1$ (today) for heavy neutrinos as a function of $m_{\nu_{\mathrm{h}}}$ for the same $x_b$ value. 
Note again that the production increases as $r_b$ increases.
We have observed that for the chosen value of $x_b$ only masses of the order $m_{\nu_{\mathrm{h}}}=10^9$GeV give $\Omega_{\nu_{\mathrm{h}}}\lesssim 10^{-3}$ today.

Despite no analytical solution for $\beta_{\bar{k}}$ can be obtained, we can numerically integrate it for all values of $\bar{k}$ in order to obtain $n$ and $\Delta \rho$, Eqs.~\eqref{n_} and~\eqref{rho_}, respectively, in terms of the definition $x=a_0/a$ and the parameters~\eqref{parameters}. 
In the calculation of $n$, Eq.~\eqref{n_}, we computed the integral numerically. The expression outside the integral is inversely proportional to $\eta_b^3 a^3$, which gives $n\propto (x_b/10^{30}{\rm{cm}})^3 x^3$. 

In the calculation of $\Delta \rho$, there is a also a dependence on the frequency $\omega_{\bar{k}}=\sqrt{\bar{k}^2+r_b^2x_b^2/x^2}$. As we want to compare with the energy density today, we made $x=1$ (for which $\omega_{\bar{k}}=\sqrt{{\bar{k}}^2+r_b^2x_b^2}$) in order to obtain the density parameter $\Omega_0=\Delta\rho_0/\rho_{\mathrm{crit0}}$ today. 
\begin{table}[htb!]
	\renewcommand*{\arraystretch}{1.1}
	\begin{center}
		\begin{tabular}{|c|c|c|c|}
			\hline
			Particles & $n_0\  [\mathrm{cm}^{-3}]$ & $\Delta\rho_0\ [\mathrm{g/cm}^{3}]$ & $\Omega_0$ \\ \hline
			Neutrinos   & $4\times 10^{-42}$ & $5\times 10^{-76}$ & $5\times 10^{-47}$\\ \hline
			Neutrons     & $3\times 10^{-27}$ & $4\times 10^{-51}$ & $4\times 10^{-22}$\\ \hline
		\end{tabular}
		\caption{Particle density, energy density and density parameters for neutrinos and neutrons today for both $x_b=10^{24}$ e $x_b=10^{30}$.}
	\end{center}
\end{table}
\begin{table}[htb!]
	\renewcommand*{\arraystretch}{1.1}
	\begin{center}
		\begin{tabular}{|c|c|c|c|}
			\hline
			Masses [GeV] & $n_0\  [\mathrm{cm}^{-3}]$ & $\Delta\rho_0\ [\mathrm{g/cm}^{3}]$ & $\Omega_0$ \\ \hline
			$10^0$             & $3\times 10^{-27}$ & $5\times 10^{-51}$ & $5\times 10^{-22}$\\ \hline
			$10^3$             & $1\times 10^{-22}$ & $2\times 10^{-43}$ & $2\times 10^{-14}$\\ \hline
			$10^6$             & $3\times 10^{-18}$ & $5\times 10^{-36}$ & $5\times 10^{-7}$\\ \hline
			$10^9$             & $1\times 10^{-13}$ & $2\times 10^{-28}$ & $2\times 10^{1}$\\ \hline
		\end{tabular}
		\caption{Particle density, energy density and density parameters for heavier neutrinos today for $x_b=10^{30}$.}
	\end{center}
\end{table}

The approximate results for $n_0$, $\Delta\rho_0$ and $\Omega_0$ for neutrinos and neutrons are shown in Table~2, whereas for heavy neutrinos the results are shown in Table~3.

The density parameter results show that only neutral particles with very large masses can produced in a relevant amount to be compared to the current critical  \mbox{density.}

\section{Conclusion}
\label{sec5}

We have calculated fermion production in cosmological models with a quantum bounce. The background model contains radiation and dust fluids, the bounce is caused by quantum effects, and its depth is a free parameter. The masses of the fermions were in the range $0.1-10^{9}$eV, from neutrinos to have fermions outside the Standard Model. The bounce depth was parametrized by the quantity $x_b=a_0/a_b$, where $a_0$ and $a_b$ are the scale factors today and at the bounce, respectively. The fermion production depends only on the ratio between the curvature scale at the bounce and the Compton wavelength of the particle, and increases as this ratio increases. This has been verified numerically. Also, fermion production does not depend on the presence of the dust fluid because it is only important when the universe is very large, where particle production is mild. This was also verified numerically.

The results show that fermion production is very small for masses below $10^9$GeV. However, for masses of this order, fermion production can be significant, with possible physical effects, like back-reaction and consequences for baryogenesis. Hence, for any physically relevant fermion production, also for a relevant of baryogenesis, either other couplings between the fermion fields and the gravitational field beyond the minimally coupling considered here should be considered, or heavier fermions beyond the Standard Model should be examined. These are subjects that we will investigate in future works. We will also study the production of charged fermions in such models.

\color{black}
\begin{acknowledgments}
	We thank Sebasti\~ao Dias for useful discussions.
    We would like to thank CNPq of Brazil for financial support.
\end{acknowledgments}

\appendix

\section{Infrared and ultraviolet limit for $\beta_{k}$}
\label{appendix:A}

The asymptotic behaviors for the Bogoliubov coefficient $\beta_{k}$, both for large (ultraviolet) and small (infrared) frequencies, can be solved analytically given equation (\ref{z-equation}) for $\mathbf{z'}$. We have:

\begin{widetext}
\begin{equation}\label{z-equation2}
{\mathbf{z}}' = \begin{pmatrix} {\alpha_k}' \\ {\beta_k}' \end{pmatrix} = \frac{ma'k}{\omega^2}\begin{pmatrix} 0 & - \exp{(-2i\int \omega d\eta)} \\ \exp{(2 i\int \omega d\eta)}& 0 \end{pmatrix} \begin{pmatrix} \alpha_k\\ \beta_k\end{pmatrix}. 
\end{equation}
\end{widetext}

We can rewrite this differential equation in cosmic time $dt = ad\eta$ using the definitions

\begin{align}\label{eq_z_uv_t}
\tilde{\omega} &\equiv \frac{\omega}{a}, \\
\mathrm{g} &= \frac{m {a}' k}{a^2\tilde{\omega}^2} = \frac{m H k}{\tilde{\omega}^2}; \; \quad \theta = \int \tilde{\omega}(t) dt; \\
\mathrm{M} &= \begin{pmatrix} 0 & - \exp(-2i\theta) \\ \exp(2 i\theta)& 0 \end{pmatrix},
\end{align}
yielding

\begin{equation}
\label{zAp}
\dot{\mathbf{z}} = \frac{\mathrm{g}}{a} \mathrm{M} \mathbf{z} \equiv \mathrm{\tilde{g}} \mathrm{M} \mathbf{z}.
\end{equation}

\subsection{The infrared limit}

In the infrared limit, $k_{\rm{phys}}\equiv k/a \rightarrow 0$, one has,

\begin{equation}
\label{exp1A}
\mathrm{\tilde{g}} \approx \frac{Hk_{\rm{phys}}}{m} + O(k_{\rm{phys}}^3).
\end{equation}

\begin{equation}
\label{exp2A}
\theta \approx \int (m + O(k_{\rm{phys}}^2))dt.
\end{equation}
Hence,

\begin{equation}\label{z-equation2A}
\dot{{\mathbf{z}}} = \begin{pmatrix} \dot{{\alpha_k}} \\ \dot{{\beta_k}}\end{pmatrix}\approx \frac{Hk_{\rm{phys}}}{m}\begin{pmatrix} 0 & - e^{(-2im\Delta t)} \\ e^{(2 im\Delta t)}& 0 \end{pmatrix} \begin{pmatrix} \alpha_k\\ \beta_k\end{pmatrix} + ...\quad . 
\end{equation}
Using the Magnus approximation \cite{Magnus:1954zz} we expand the solution as

\begin{equation}\label{z-equation3A}
\begin{pmatrix} {\alpha_k}(t) \\ {\beta_k}(t)\end{pmatrix} \approx \begin{pmatrix} {\alpha_k}(t_i) \\ {\beta_k}(t_i)\end{pmatrix} +k_{\rm{phys}}\begin{pmatrix} 0 & f_1(t)\\ f_2(t)& 0 \end{pmatrix} \begin{pmatrix} \alpha_k(t_i)\\ \beta_k(t_i)\end{pmatrix} + ...\quad . 
\end{equation}
For

\begin{equation}\label{z-equation4A}
\begin{pmatrix} {\alpha_k}(t_i) \\ {\beta_k}(t_i)\end{pmatrix} =  \begin{pmatrix} 1 \\ 0 \end{pmatrix} , 
\end{equation}
one gets the result ${\beta_k}(t) = k_{\rm{phys}} f_2(t) + O(k_{\rm{phys}}^2)$.

\subsection{The ultraviolet limit}

In the ultraviolet limit, $k_{\rm{phys}}\equiv k/a \rightarrow \infty$, $\mathrm{\tilde{g}}$ is also a small parameter, $$\mathrm{\tilde{g}}\approx \frac{mH}{k_{\rm{phys}}}[1 + O(k_{\rm{phys}}^{-2})],$$ but the matrix $\mathrm{M}$ also depends on $k$. However, integrating the differential equation for $\mathbf{\dot{z}}$ in Eq.~(\ref{eq_z_uv_t}), we have:

\begin{align}
\mathbf{z} &= \mathbf{z}(t_{i}) + \int^{t} \tilde{g}_{1} \mathrm{M}_{1} \mathbf{z}_{1} dt_{1}\\
&= \mathbf{z}(t_{i}) + \int^{t} \tilde{g}_{1} \mathrm{M}_{1} \left[ \mathbf{z}(t_{i}) + \int^{t_{1}} \tilde{g}_{2} \mathrm{M}_{2} \mathbf{z}_{2} dt_{2} \right] dt_{1} \\
&= \mathbf{z}(t_{i}) + \mathbf{z}(t_{i})\int^{t} \tilde{g}_{1} \mathrm{M}_{1} dt_{1} \left[ 1 + \int^{t_{1}} \tilde{g}_{2} \mathrm{M}_{2} dt_{2} \right] + \ldots \label{z_uv_2_ult}
\end{align}

We can now turn to the analysis of the integrals present in Eq.~(\ref{z_uv_2_ult}),

\begin{equation}
\int^{t} \tilde{g}_{1} \mathrm{M}_{1} dt_{1} = \begin{pmatrix} 0 & - \int^{t} \tilde{g}_{1} e^{(-2i\theta)}dt_{1} \\ \int^{t} \tilde{g}_{1}e^{(2 i\theta)} dt_{1}& 0 \end{pmatrix} .
\end{equation}

Integrating the nonzero terms, we get:

\begin{align}\label{integral_M_z}
\int^{t} &\tilde{g}_{1} \exp(\pm 2i\theta)dt_{1} = \int^{t} \tilde{g}_{1} \frac{\mp 2i\tilde{\omega}_{1}}{\mp 2i\tilde{\omega}_{1}}\exp(\mp2i\theta)dt_{1} \\
&= \frac{\tilde{g}_{1}\exp(\mp2i\theta)}{\mp 2i\tilde{\omega}_{1}} - \int^{t}\left(\frac{\tilde{g}_{1}}{\mp 2i\tilde{\omega}_{1}} \right)^{.}\exp(\mp2i\theta) dt_{1} \\
&= \frac{\tilde{g}_{1}\exp(\mp2i\theta)}{\mp 2i\tilde{\omega}_{1}} - \left(\frac{\tilde{g}_{1}}{\mp 2i\tilde{\omega}_{1}} \right)^{.}\frac{\exp(\mp2i\theta)}{{\mp 2i\tilde{\omega}_{1}}} + ... \label{integral_M_z3}
\end{align}

Comparing the magnitude of the first two terms above, one gets

\begin{align}
\frac{\left(\tilde{g}_{1}/\tilde{\omega}_{1} \right)^{.}}{\tilde{g}} 
= \frac{1}{\tilde{\omega}^3}\left[ \left(\frac{\dot{H}}{H} + 2H\right) \left(\frac{k^2}{a^2}\right) + \left(\frac{\dot{H}}{H} + 2H\right) \left(m^2\right)  \right], \label{razao_UV}
\end{align}
which, in the UV limit $k_{\rm{phys}} = k/a \sim \infty$, yields

\begin{equation}
\lim_{UV}\frac{\left(\tilde{g}_{1}/\tilde{\omega}_{1} \right)^{.}}{\tilde{g}} \approx \frac{\left(\dot{H}/H + 2H\right)}{k_{\rm{phys}}}\approx 0 .
\end{equation}

Therefore, in the UV limit, keeping only the first term, one gets for $\mathbf{z}$

\begin{equation}
\mathbf{z} \approx \mathbf{z}(t_{i}) + \begin{pmatrix} 0 &\frac{\tilde{g}_{1}\exp(-2i\theta)}{- 2i\tilde{\omega}_{1}} \\ \frac{\tilde{g}_{1}\exp(2i\theta)}{ 2i\tilde{\omega}_{1}}& 0 \end{pmatrix} \mathbf{z}(t_{i}).
\end{equation}
Hence,

\begin{equation}
\mathbf{z} =  \begin{pmatrix} \alpha_{k} \\ \beta_{k} \end{pmatrix}  = \begin{pmatrix} 1 \\ 0 \end{pmatrix} +  \begin{pmatrix} 0 \\ \frac{\tilde{g}_{1}\exp(-2i\theta)}{- 2i\tilde{\omega}_{1}}  \end{pmatrix},
\end{equation}
and the behavior of $\beta_{k}$ in the UV limit reads

\begin{align}
\beta_{k} \sim \frac{\tilde{g}_{1}\exp(-2i\theta)}{- 2i\tilde{\omega}_{1}} =\frac{-\exp(2i\theta)}{2i}\frac{mHk}{\tilde{\omega}^3a} \sim \frac{1}{k_{{\rm{phys}}}^2} .
\end{align}

Therefore, the Bogoliubov coefficient $\beta_{k}$ decays as $k_{\rm{phys}}^2$ for large $k_{\rm{phys}}$.





\begin{thebibliography}{99}
\bibitem{Kolb:1990vq} 
  E.~W.~Kolb and M.~S.~Turner,
  \textit{The Early Universe}
  (Avalon Publishing, 1994)
  Front.\ Phys.\  {\bf 69}, 1 (1990).

\bibitem{Lyth:1998xn} 
  D.~H.~Lyth and A.~Riotto,
  Phys.\ Rept.\  {\bf 314}, 1 (1999)



\bibitem{Riotto} 
G.~F.~Giudice, M.~Peloso, A.~Riotto and I.~Tkachev,
JHEP {\bf 9908}, 014 (1999)

\bibitem{Peloso:2000hy} 
M.~Peloso and L.~Sorbo,
JHEP {\bf 0005}, 016 (2000)




\bibitem{Mukhanov:2005sc} 
  V.~Mukhanov,
  \textit{Physical Foundations of Cosmology}
  (Cambridge University Press, Oxford, 2005)


\bibitem{Finelli:2001sr} 
  F.~Finelli and R.~Brandenberger,
  Phys.\ Rev.\ D {\bf 65}, 103522 (2002)

  
\bibitem{Peter:2006hx} 
  P.~Peter, E.~J.~C.~Pinho and N.~Pinto-Neto,
  Phys.\ Rev.\ D {\bf 75}, 023516 (2007)


\bibitem{Peter:2008qz} 
  P.~Peter and N.~Pinto-Neto,
  Phys.\ Rev.\ D {\bf 78}, 063506 (2008)
  


\bibitem{Vitenti:2011yc} 
  S.~D.~P.~Vitenti and N.~Pinto-Neto,
  Phys.\ Rev.\ D {\bf 85}, 023524 (2012)

\bibitem{Peter:2015zaa} 
  P.~Peter, N.~Pinto-Neto and S.~D.~P.~Vitenti,
  Phys.\ Rev.\ D {\bf 93}, no. 2, 023520 (2016)
  
\bibitem{Celani:2016cwm} 
D.~C.~F.~Celani, N.~Pinto-Neto and S.~D.~P.~Vitenti,
Phys.\ Rev.\ D {\bf 95}, no. 2, 023523 (2017)

\bibitem{Wands:1998yp} 
  D.~Wands,
  Phys.\ Rev.\ D {\bf 60}, 023507 (1999)

\bibitem{Bacalhau:2017hja} 
A.~P.~Bacalhau, N.~Pinto-Neto and S.~Dias Pinto Vitenti,
Phys.\ Rev.\ D {\bf 97}, no. 8, 083517 (2018)

\bibitem{Quintin:2015rta} 
  J.~Quintin, Z.~Sherkatghanad, Y.~F.~Cai and R.~H.~Brandenberger,
  Phys.\ Rev.\ D {\bf 92}, no. 6, 063532 (2015)

\bibitem{Ashtekar1} 
  A.~Ashtekar,
  in \textit{Quantum Gravity and Quantum Cosmology}
  (Springer-Verlag, Berlin, 2013), pp. 31-56.
  
\bibitem{Ashtekar2} 
  A.~Ashtekar,
  in \textit{General Relativity}, Cosmology and Astrophysics
  (Springer, New York, 2014), pp. 323-347.
  
\bibitem{Pinto-Neto:2013toa} 
  N.~Pinto-Neto and J.~C.~Fabris,
  Class.\ Quant.\ Grav.\  {\bf 30}, 143001 (2013)
  




\bibitem{HollandBook} 
  P.~R.~Holland,
  \textit{The Quantum Theory of Motion}
  (Cambridge University Press, Oxford, 1993),



  
\bibitem{AcaciodeBarros:1997gy} 
J.~Acacio de Barros, N.~Pinto-Neto and M.~A.~Sagioro-Leal,
Phys.\ Lett.\ A {\bf 241}, 229 (1998)

\bibitem{Alvarenga:2001nm} 
F.~G.~Alvarenga, J.~C.~Fabris, N.~A.~Lemos and G.~A.~Monerat,
Gen.\ Rel.\ Grav.\  {\bf 34}, 651 (2002)

\bibitem{PintoNeto:2005gx} 
N.~Pinto-Neto, E.~S.~Santini and F.~T.~Falciano,
Phys.\ Lett.\ A {\bf 344}, 131 (2005)

\bibitem{RiottoTrodden} 
A.~Riotto and M.~Trodden,
Ann.\ Rev.\ Nucl.\ Part.\ Sci.\  {\bf 49}, 35 (1999)

\bibitem{Antunes:2016efv} 
  V.~Antunes, I.~Bediaga and M.~Novello,
  arXiv:1611.07802

\bibitem{Davoudiasl:2004gf} 
  H.~Davoudiasl, R.~Kitano, G.~D.~Kribs, H.~Murayama and P.~J.~Steinhardt,
  Phys.\ Rev.\ Lett.\  {\bf 93}, 201301 (2004)


\bibitem{Parker:1971pt} 
  L.~Parker,
  Phys.\ Rev.\ D {\bf 3}, 346 (1971)
  Erratum: [Phys.\ Rev.\ D {\bf 3}, 2546 (1971)].
  doi:10.1103/PhysRevD.3.346, 10.1103/PhysRevD.3.2546


\bibitem{Kinney:2005in} 
W.~H.~Kinney and A.~Riotto,
JCAP {\bf 0603}, 011 (2006)

\bibitem{BirrelDaviesBook} 
N.~D.~Birrell and P.~C.~W.~Davies,
\textit{Quantum Fields in Curved Space},
Cambridge Monographs on Mathematical Physics
(Cambridge University Press, 1982)

\bibitem{ParkerTomsBook} 
L.~E.~Parker and D.~Toms,
\textit{Quantum Field Theory in Curved Spacetime : Quantized Field and Gravity},
Cambridge Monographs on Mathematical Physics
(Cambridge University Press, 2009)


\bibitem{WaldBook} 
R.~M.~Wald,
\textit{Quantum Field Theory in Curved Space-Time and Black Hole Thermodynamics},
Chicago Lectures in Physics
(University of Chicago Press, 1994)

\bibitem{Chung:2003wn} 
D.~J.~H.~Chung, A.~Notari and A.~Riotto,
JCAP {\bf 0310}, 012 (2003)

\bibitem{Quintin:2014oea} 
J.~Quintin, Y.~F.~Cai and R.~H.~Brandenberger,
Phys.\ Rev.\ D {\bf 90}, no. 6, 063507 (2014)

\bibitem{Haro:2015zda} 
J.~Haro and E.~Elizalde,
JCAP {\bf 1510}, no. 10, 028 (2015)

\bibitem{Tavakoli:2014mra} 
Y.~Tavakoli and J.~C.~Fabris,
Int.\ J.\ Mod.\ Phys.\ D {\bf 24}, no. 08, 1550062 (2015)

\bibitem{Hipolito-Ricaldi:2016kqq} 
W.~S.~Hipolito-Ricaldi, R.~Brandenberger, E.~G.~M.~Ferreira and L.~L.~Graef,
JCAP {\bf 1611}, no. 11, 024 (2016)

\bibitem{Tsujikawa:2001ud} 
S.~Tsujikawa and H.~Yajima,
Phys.\ Rev.\ D {\bf 64}, 023519 (2001)

\bibitem{Linde:94} 
L.~Kofman, A.~D.~Linde and A.~A.~Starobinsky,
Phys.\ Rev.\ Lett.\  {\bf 73}, 3195 (1994)

\bibitem{Linde:97} 
L.~Kofman, A.~D.~Linde and A.~A.~Starobinsky,
Phys.\ Rev.\ D {\bf 56}, 3258 (1997)

\bibitem{Amin_Reheating} 
M.~A.~Amin, M.~P.~Hertzberg, D.~I.~Kaiser and J.~Karouby,
Int.\ J.\ Mod.\ Phys.\ D {\bf 24}, 1530003 (2014)

\bibitem{Brandenberger_Reheating} 
R.~Allahverdi, R.~Brandenberger, F.~Y.~Cyr-Racine and A.~Mazumdar,
Ann.\ Rev.\ Nucl.\ Part.\ Sci.\  {\bf 60}, 27 (2010)

\bibitem{GellMann:1980vs} 
M.~Gell-Mann, P.~Ramond and R.~Slansky,
Conf.\ Proc.\ C {\bf 790927}, 315 (1979)

\bibitem{Neutrino_mass} 
R.~N.~Mohapatra and G.~Senjanovic,
Phys.\ Rev.\ Lett.\  {\bf 44}, 912 (1980).

\bibitem{Ballesteros:2016xej} 
G.~Ballesteros, J.~Redondo, A.~Ringwald and C.~Tamarit,
JCAP {\bf 1708}, no. 08, 001 (2017)


\bibitem{gradshteyn} 
  I.~S.~Gradshteyn, I.~M.~Ryzhik,
  \textit{Table of Integrals, Series, and Products}
  (Academic Press, 2014)



\bibitem{Mertens:2016ihw} 
S.~Mertens,
J.\ Phys.\ Conf.\ Ser.\  {\bf 718}, no. 2, 022013 (2016)

\bibitem{Magnus:1954zz}
  W.~Magnus,
  Commun.\ Pure Appl.\ Math.\  {\bf 7}, 649 (1954)

\end{thebibliography}

\end{document}